# NutriFD: Proving the medicinal value of food nutrition based on food-disease association and treatment networks

## Abstract

Wanting Su[1,2,3], Dongwei Liu[4], Feng Tan[2], Lun Hu[1], Pengwei Hu[1,2*]

There is rising evidence of the health benefit associated with specific dietary interventions. Current food-disease databases focus on associations and treatment relationships but haven't provided a reasonable assessment of the strength of the relationship, and lack of attention on food nutrition. There is an unmet need for a large database that can guide dietary therapy. We fill the gap by developing NutriFD, a scoring network based on associations and therapeutic relationships between foods and diseases. NutriFD integrates 9 databases including foods, nutrients, diseases, genes, miRNAs, compounds, disease ontology and their relationships. To our best knowledge, this database is the only one that can score the associations and therapeutic relationships of everyday foods and diseases by weighting inference scores of food compounds to diseases. In addition, NutriFD demonstrates the predictive nature of nutrients on the therapeutic relationships between foods and diseases through machine learning models, laying the foundation for a mechanistic understanding of food therapy.

***Keywords***：Food Dietary, Nutrition, Food-disease Networks, Machine Learning Models

[1]Xinjiang Technical Institute of Physics and Chemistry, Chinese Academy of Sciences, Urumqi, China. [2]Merck KGaA, Science and Technology Office. [3]The University of Manchester, Division of Informatics, Imaging & Data Sciences. [4]ByteDance (Nanjing) Technology R&D Co., Ltd. ✉email: hpw@ms.xjb.ac.cn

# Introduction

The Global Burden of Disease (GBD) 2017 Diet Collaborators used observational data and short-term trials of intermediate outcomes to establish causal associations between individual dietary components and death and disease (1). Nutrients are essential for the proper functioning of the human body as they provide energy for activity, materials for growth and repair, and maintain a healthy immune system (2). Moreover, nutritional deficiencies are major risk factors for many chronic diseases and conditions, such as cardiovascular disease, stroke, type 2 diabetes, obesity, cancer, and others. Hence, The 2020-2030 NIH Nutrition Research Strategic Plan is focused on advancing precision nutrition to answer the key question: "How can we improve the use of food as medicine?" (3).

Studies have shown that naturally occurring bioactive compounds, including those found in turmeric, broccoli sprouts, berries, propolis, and other foods, have the potential to be used as nutritional therapies for chronic kidney disease by regulating the expression of pro-inflammatory transcription factors and inflammatory cytokines (4). Specifically, nutrients such as catechins, sulforaphane, and curcumin that stimulate the cell-protective transcription factor nuclear factor erythroid 2-related factor 2 (NRF2) improve mitochondrial function and reduce aging through induction of the anti-aging pathway (5).

Nutrients are not the only compounds found in food that can impact disease development and progression. Non-nutrient bioactive compounds, such as resveratrol found in nuts and red wine, have antioxidant, anti-thrombotic, anti-inflammatory properties, and can even inhibit carcinogenesis (6). Therefore, it is crucial to consider both nutrients and bioactive compounds when researchers examine the relationship between food and disease.

Traditional methods for studying the connection between food and disease often rely on time-consuming long-term experiments (7,8). However, with the vast amount of biomedical data available, it is now possible to build links between diseases and foods through compounds using data-driven approaches. This study aims to develop a dataset for exploring the relationship between foods and diseases from the perspective of nutrients. This dataset can be used to guide AI dietary therapy for regulating chronic disease courses through daily diet. In addition, it can be used to investigate potential relationships between foods and diseases, providing directions for future research for biologists.

Currently, food-disease datasets fall into two primary categories. The first category comprises associations between plant foods and diseases, such as NutriChem (9) and FooDisNET (10). NutriChem contains 6242 text-mining connections between 1066 plant foods and 751 diseases, while FooDisNET links 6329 plant foods and 18689 diseases. The second type is traditional Chinese medicine (TCM) datasets (11–13), which examine the therapeutic relationship between herbs and diseases. For instance, TCMID 2.0 (11) includes 1356 TCMs and their associations with 842 diseases. Nonetheless, since most herbs are not part of daily human dietary intake, TCM datasets are not suitable for this study.

The accuracy and validity of food-disease databases depend largely on the methodology used to link foods and diseases. One approach involves extracting food-disease associations directly through text mining. For instance, the MAPS database (14) extracted data on over 500 medicinal plants from PubMed papers.

The second method is to connect foods and diseases by agent, such as compound. NutriChem was reportedly the first to link foods and diseases through compounds which were represented by simplified molecular-input line-entry system (SMILES). To obtain further in-depth information on foods and diseases, NutriChem 2.0 considered the interaction of overlapping targets between food compounds and drugs (15). While SMILES is widely used for storage and interchange of chemical structures (16), no standard exists to generate a canonical SMILES string (17). Some matching bias may be caused. Nevertheless, the InChI is a more precise representation of compound, and aims to provide a unique, or canonical identifier for chemical structures (17,18). FooDisNet (10) established compound connections between plant foods and disease targets through InChiKey, a compression of InChi, which improved the accuracy of integrating compounds. In addition to "compound-target" linkage method, ABCkb (19) established such associations by "compound-gene" linkages.

Regarding the assessment of the relationship between foods and diseases, FooDisNet evaluated associations between plant foods and diseases by developing four scoring strategies (10). These strategies included the prevalence of food, the rarity of compounds, specificity of proteins, and universality of the disorder. However, this approach reduced the usability and objectivity of the data. While NutriChem provided information on 'preventive' and 'promoting' associations between foods and diseases, it did not propose a method for assessing such relationships. Previous studies investigating relationships between foods and diseases have not utilized a standardized method for quantitative analysis.

Although several food-disease databases have been constructed, most of them are limited in compounds integration accuracy and scoring method objectivity. Moreover, lack of attention on food nutrition has existed as a problem for many years since the relation exploration between food nutrients and diseases is a major contributor to food therapy. In addition, what is not yet clear is the assessment of the treatment relation from foods to diseases.

To address these gaps, we have created NutriFD, which includes both NutriFDA and NutriFDT. NutriFDA focuses on scoring associations between foods and diseases, while NutriFDT focuses on scoring treatment relations. This resource enables researchers to browse and analyze food-disease associations and treatment relations from a nutrient-based perspective. The current version of NutriFDA consists of 3237640 connections between 591 foods and 2208 diseases, while NutriFDT consists of 170531 connections between 591 foods and 671 diseases. NutriFD provides nutrient features and disease similarity features based on disease-mirna associations, disease-gene associations, and disease ontology. To prove the predictive nature of nutrients on food dietary, we developed multi-classification tasks of NutriFDT and NutriFDA by machine learning models. We anticipate that this database will be widely utilized and will promote new avenues for research in the field of food therapy.

# Methods

## Overview

The process of constructing NutriFD was conducted in six stages (see Figure 1). The first five steps involved dataset creation, while the final step was devoted to validating the quality of datasets. We began by acquiring a food-nutrition dataset from USDA (20). The second step involved constructing two types of food-disease relationships: scoring associations and scoring treatment relations. Additionally, to generate disease characteristics, we developed three disease similarity datasets based on disease-mirna associations, disease-gene associations, and disease ontology. After consolidating the consensus foods and diseases from these datasets, we con-structed two food-disease networks based on different food-disease relationships: NutriFDA considering associations and NutriFDT considering treatment relations. The structure of the networks is characterized by food and disease nodes, the connections between them, food nu-trition characteristics, and disease similarity characteristics. Ultimately, the generalization per-formance of the networks was evaluated by developing multi-classification tasks for NutriFDA and NutriFDT, employing machine learning models. The overview of food-disease network is presented in Figure 2.

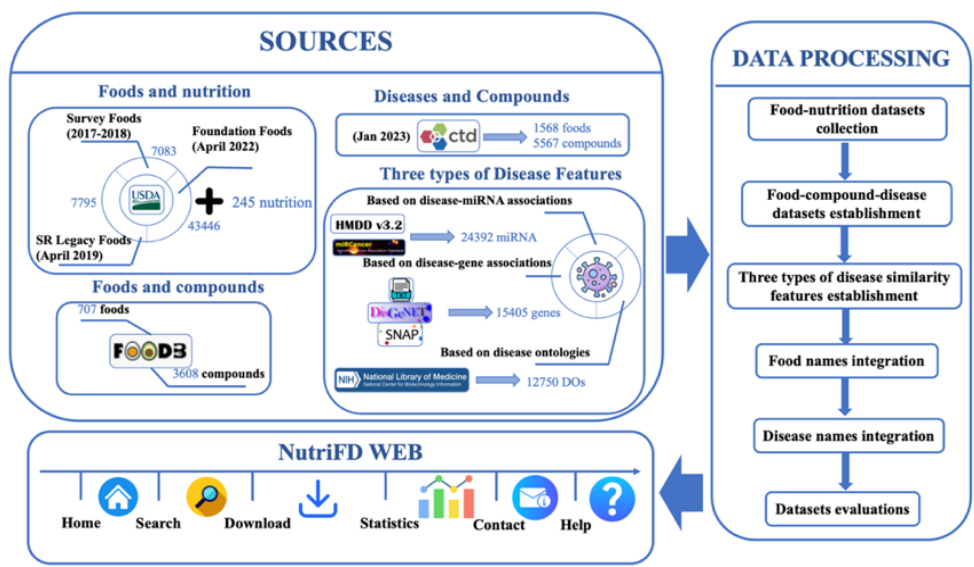

**Figure 1 Process of NutriFD establishment**

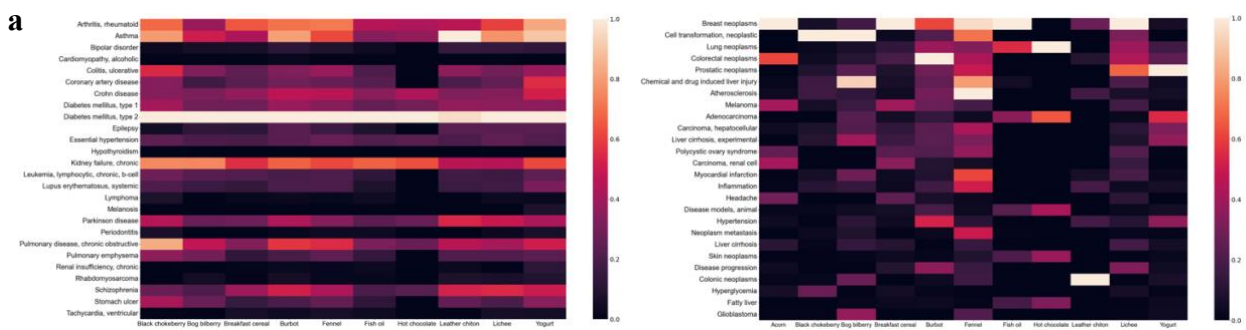

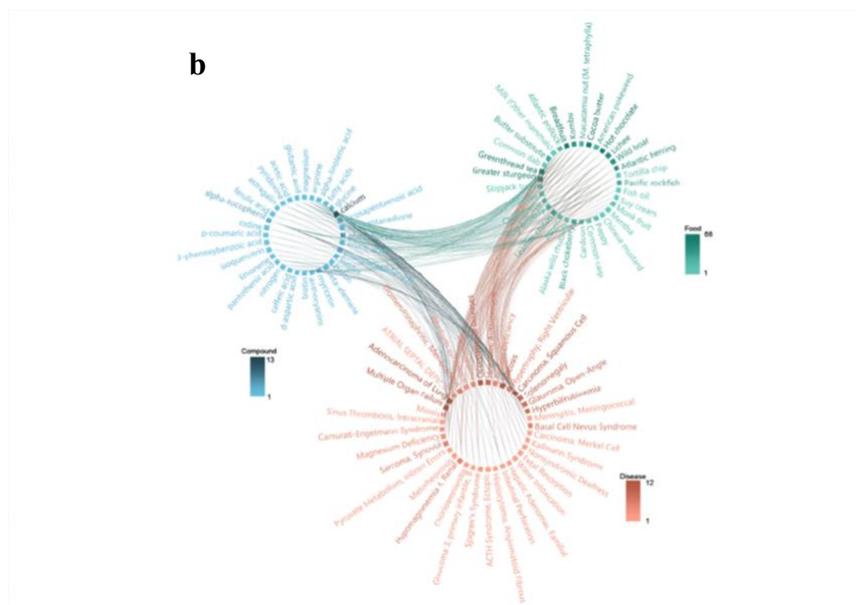

**Figure 2 Food-Disease network overview.** a. Heatmaps show normalized scores of the associations and treatments between 11 selected foods and a subset of diseases. The above heatmap represents the association scores while the below heatmap represents the treatment scores. b. A food-compound-disease network was constructed to demonstrate the scoring system for evaluating the relationships between foods and diseases via compounds. Specifically, a subset of foods and diseases were selected, and their scores were calculated based on the compounds connecting them.

## Settings & Assumptions

To establish NutriFDA and NutriFDT, we set the below assumptions:

- InChiKey and CASRN to represent compounds for connecting food-compound and compound-disease are assumed to be unique. Although there is a possibility of two different compounds having the same InChiKey, this is extremely unlikely (21).

- We introduced association and treatment inference scores of compounds to diseases from CTD (22,23). They performed a weighted scoring inference of relationships between compounds and diseases through chemical-gene-disease networks by evaluating the hypergeometric clustering coefficient (24), as well as two common neighbor statistics (25). This scoring inference allowed us to determine the degree of association between compounds and diseases.

- When integrating food names from FooDB (26) and USDA, we focused on integrating everyday foods with high unprocessed level. For unmatched FooDB foods, we used partial and handy matching methods to find the best match in the USDA dataset, as shown in the Figure 3(c), assuming that such FooDB foods are equivalent to the matched USDA foods.

## Materials and methods

a

b

c

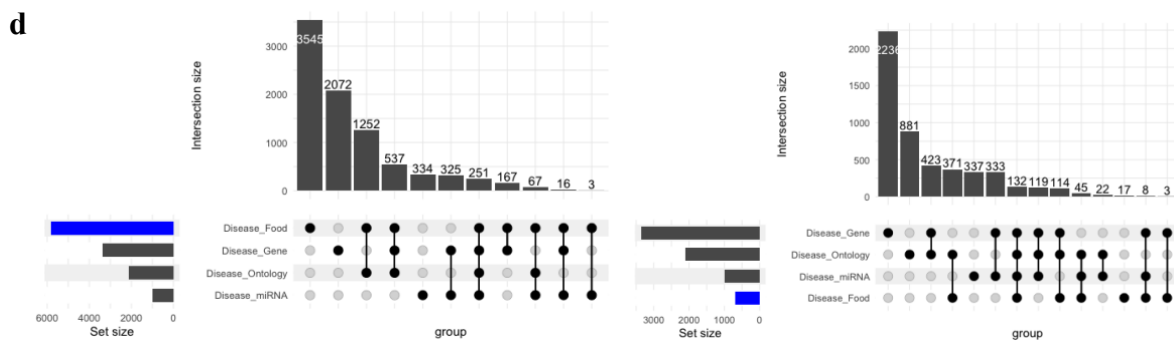

**Figure 3 The workflow of dataset construction.** a. The figure represents the method to calculate adjacent matrix between foods and diseases. It displays a schematic representation of the variables considered in the study. Specifically, it shows the designation of $D_i$ for the i-th disease, $C_j$ for the j-th compound, and $F_k$ for the k-th food item. b. The figure represents methods to calculate three types of disease features. Before calculating the semantic features of diseases using disease ontology, it is important to note that we first took the intersection of diseases with ontology and those related to foods. This resulted in a total of 2109 diseases for semantic similarity calculation. c. The figure represents the process of integrating food names. The term "unprocessed foods" refers to food items listed by the USDA with names containing "100%", "raw" and "NFS". The phrase "core foods" refers to the part of the food name before the first comma. The study uses different methods for matching food items from the FooDB and USDA databases. "Full match" requires that every word in the FooDB food name is present in the USDA food name. "Partial and manual match" accepts matches where at least one word from the FooDB food name is present in the USDA food name and filters manually the best match. "Homology de-duplication" retains the latest record and calculates the median nutrition amount for foods with similar names. "Heterogeneous de-duplication" selects the food item with the highest priority according to the source priority in Table 1. These definitions and methods are used consistently throughout the paper to ensure clarity and consistency in the analysis of the data. d. The left figure represents intersection of disease names from different sources in NutriFDA while the right figure represents that in NutriFDT. The blue highlighted portion of the figure represents the diseases that are related to food.

## (1) Food-nutrition collection and criteria

Food-nutrition datasets were collected from USDA by limiting the food type to "Foundation Foods", "SR Legacy Foods", and "Survey Foods", which contain everyday foods with high unprocessed levels.

## (2) Food-compound-disease datasets building criteria

Compound CASRNs were achieved by FooDB, CTD and PubChem API. Compound InChiKeys were achieved by PubChem API and ChemSpider API. The search approach was followed by manual curation. In the end, we obtained InChiKeys for 90.3% and CASRNs for 77.5% of the compounds. The relationships between foods and diseases were linked by 848 compounds, representing 22.9% of food compounds.

To calculate the inference scores between foods and diseases, a new method was proposed as indicated by Figure 3(a).

Inference scores between compounds and diseases were obtained from CTD, which were then combined with the food compound content to calculate the inference scores between foods and diseases. For the treatment relations between foods and diseases, therapeutic evidence data from CTD were selected. Using the method proposed in Figure 3(a), the resulting association matrix had a shape of 684*5840, while the inference score matrix for treatment relations had a shape of 684*691.

**(3) Three types of disease similarity features**

To consider as much disease information as possible, we incorporated disease similarity features obtained from disease-gene, disease-mirna, and disease ontology. These features were utilized separately in our model validation. Specifically, we constructed models using each of these features to ensure comprehensive analysis of the data. To create three types of disease similarity datasets, we utilized disease Gaussian similarity for disease-miRNA associations and disease-gene associations, and disease semantic similarity for disease ontology. The calculation process is represented in Figure 3(b).

**1) Disease semantic similarity (27)**

Diseases can be classified by Medical Subject Headings (MeSH) disease descriptors in a strict system. In the NCBI database (28), each disease is represented by a Directed Acyclic Graph (DAG) where diseases are nodes and edges connect the parent and child nodes. Given any disease $d$, the corresponding directed acyclic graph is denoted as $DAG_d = (D, C_d, P_d)$, where $C_d$ is the child nodes of $d$ and $P_d$ is the parent nodes of $d$. Therefore, any a disease term $d$ in $DAG_d$, the semantic characteristic is defined as,

$$\begin{cases} D_d(t) = 1 \\ D_d(t) = max\{\Delta^* D_d(t') \mid t' \in \text{ children of } t\} \text{ if } t \neq d \end{cases} \quad (1)$$

In simpler terms, the decay factor ($\Delta$) determines the extent to which ancestor diseases contribute to the semantic value of disease $d$. As the distance between a disease and disease $d$ increases, the contribution of its ancestor diseases decreases. Disease $d$ is in the 0th layer of the DAG, which makes it the most specific disease term with a contribution of 1 to the semantic value of disease $d$. Diseases in the 1st layer are considered more general, so their contribution is multiplied by the decay factor which is 0.5. This formula differentiates the semantic contribution of diseases in different layers to the semantic value of disease $d$.

The semantic value of a disease d is calculated as the sum of the semantic contribution values of all diseases in its corresponding DAG, as defined below,

$$DV(d) = \sum_{t \in T_d} D_d(t) \quad (2)$$

Hence, for any disease $d_1$ and $d_2$, the semantic similarity between them can be defined as,

$$Sim(D_{d1}, D_{d2}) = \frac{\sum_{t \in D_{d1} \cap D_{d2}} (D_{d1}(t) + D_{d2}(t))}{\sum_{t \in D_{d1}} (D_{d1}) + \sum_{t \in D_{d2}} (D_{d2})} \quad (3)$$

**2) Disease Gaussian similarity** (29)

We computed the Gaussian interaction distribution nuclear similarity between two diseases (denoted as $d_1$ and $d_2$) in the matrix M, which represents disease features. In particular, the i-th row of D is the vector $V(d_i)$, which describes the feature of disease $i$. The similarity between $d_1$ and $d_2$ can be defined as follows,

$$sim_M(d_1, d_2) = exp\left(-\gamma_d * \| V(d_1) - V(d_2) ) \|^2\right) \quad (4)$$

where $\gamma_d$ parameter is applied to regulate the kernel bandwidth and $nd$ is the column amount of *M*, which can be defined as,

$$\gamma_d = 1 / \left(\frac{1}{nd}\sum_{i=1}^{nd} \|V(D_i)\|^2\right) \qquad (5)$$

**(4) The process of foods integration**

To unify the foods between the food-disease dataset and the food-nutrition dataset, we took into consideration the discrepancies between FooDB (26) and USDA, as shown in Figure 3(c).

To obtain USDA foods with high unprocessed levels, we ranked the unprocessed degrees of food sources based on a priority list (Table 1), since foods with the same core term in USDA may have different processing methods. In cases where a FooDB food had multiple USDA food records, we selected the USDA food with the highest unprocessed degree (Figure 3(c)). As a result, we were able to match 591 foods containing nutrition information with 487 foods containing compound information.

**Table 1 Rankings of food sources**

| Source | Data type | Description | Rank |
|---|---|---|---|
| Foundation | foundation_food | Foods from many samples | 1 |
| Foundation | sub_sample food | Food sample from sample_food | 5 |
| Foundation | market acquisition | Foods collected for chemical experiments | 3 |
| Foundation | agriculture_ acquisition | Foods collected directly from the original places | 3 |
| Foundation | sample_food | The sample food of foundation_food | 4 |
| SR_legacy | sr_legacy_food | Foods collected from USDA | 2 |

| | | National Nutrient Database | |
| | | Foods collected for consumption | |
| Survey | survey_fndds_food | of food nutrients | 6 |

**(5) The process of diseases integration**

To achieve better integration of diseases, we removed certain insignificant words such as "disease", "to", "syndrome", "disorders", "and", "of", "or" and "with" from disease names. Then, we used an exact match method to integrate disease names between diseases containing features and those related to foods. The matching result between different types of disease is shown in Figure 3(d).

Finally, 2208 diseases associated with food have a minimum of one feature type in NutriFDA, and the same was found for 671 diseases in NutriFDT.

**(6) Evaluation and Prediction**

The paper used machine learning methods to evaluate the generalization capability of NutriFD datasets. The process was divided into the following steps,

1) Independent variables were generating by connecting the nutritional characteristics of each food with the similarity characteristics of each disease.

2) Corresponding dependent variables were the corresponding value of the food-disease relationship matrix.

3) Data pre-processing was to classify the value of 0 in the dependent variable as a separate category, then group the remaining values using a binning method.

4) Machine learning models including decision tree models and random forest models were used.

5) The f1-score and AUC in the testing datasets were reported as evaluation indicators.

# Results

After the above steps, we generated NutriFDA and NutriFDT networks based on food-disease associations and treatment relations, as shown in Figure 4.

## A. Networks Study

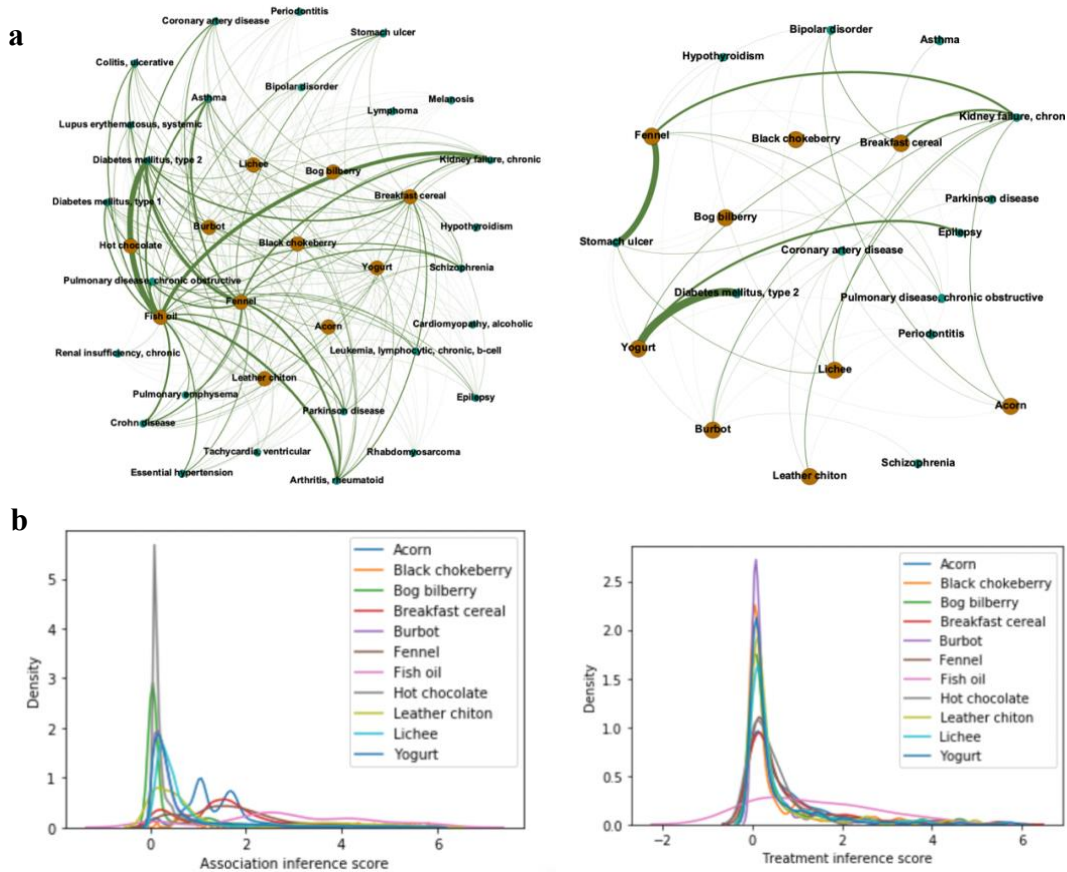

**Figure 4 Networks analysis.** a. A partial treatment network between foods and diseases is shown for NutriFDT (right) while a partial association network is shown for NutriFDA (left). Diseases and foods are represented by blue and yellow nodes, respectively. The thickness of the edges reflects the strength of the association scores and therapeutic scores. b. Comparison of the distribution of association scores and therapeutic inference scores for eleven selected foods are presented. Kernel density plots show the density of scores. Inference scores less than 6 and not equal to 0 are included.

There are fewer relationships capable of producing therapeutic effects than those capable of producing associations, as shown in Figure 4(a). The strongest association observed was between "Fish oil" and "Diabetes mellitus, type 2" (30,31), followed by "Fish oil" and "Kidney failure, chronic" (32). However, the strongest therapeutic relationship was found between "Yogurt" and "Diabetes mellitus, type 2" (33), followed by "Fennel" and "Stomach ulcer" (34). These relationships are supported by previous research indicating that these foods may have therapeutic effects on specific diseases, mediated by complex relationships between multiple compounds and targets.

When examining the plot of distributions of association inference scores from Figure 4(b). we found that most inference scores of foods were concentrated between 0 and 0.7, with the highest peak observed for fennel. However, acorn and breakfast cereal produced small peaks between 1 and 2. For fish oil, its highest peak was observed at around 2.5, indicating a relatively strong association with certain diseases. In contrast, when examining the plot of distributions of treatment inference scores from Figure 4(b) the peak of inference scores for all foods were concentrated between 0 and 0.5, with the highest peak observed for burbot. However, fish oil displayed a broader and smoother curve, indicating its potential therapeutic effects across a wide range of diseases.

These findings suggest that different foods may have varying degrees of association and therapeutic effects on different diseases. Further research is needed to fully understand the complex relationships between foods and diseases.

## B. Case Study

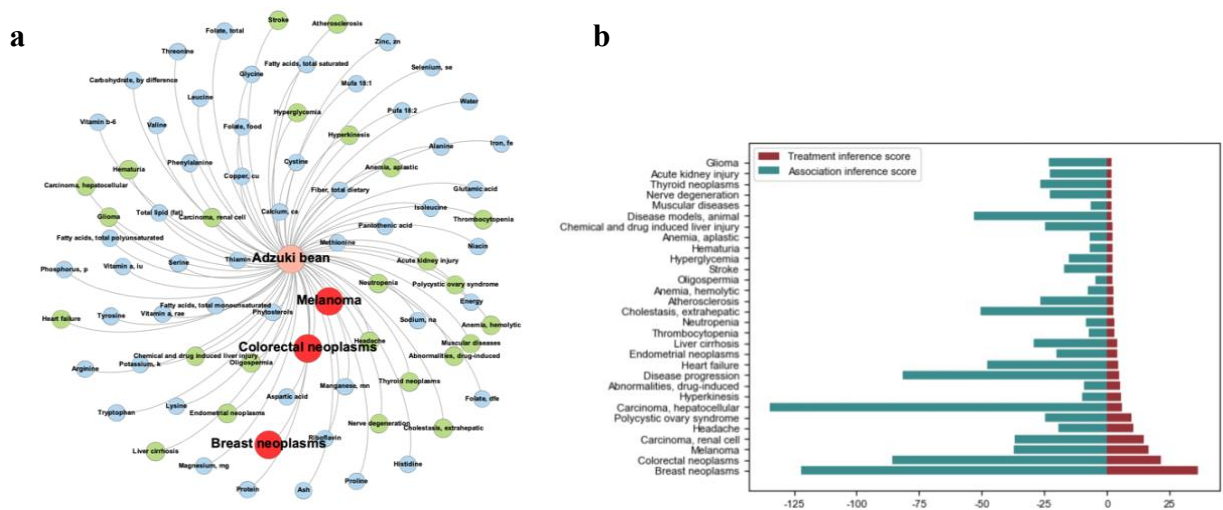

**Figure 5 Case analysis.** a. The figure represents nutrient-disease network of adzuki bean. A treatment relationship network for adzuki beans were constructed by adding its relationships with 49 nutrients and 27 diseases. Blue nodes represented nutrients, while green nodes represented diseases. The top three diseases with the strongest treated relationships by adzuki bean were highlighted in red. b. The figure represents comparing association and treatment inference scores of adzuki bean to 30 diseases. Top 30 diseases treated by adzuki bean were selected.

Adzuki beans, a type of legume commonly used in Asian cuisine and medicine, exhibited the strongest dose-dependent inhibitory effect against ovary cancer cell SK-OV-3 and breast cancer cell MCF-7 compared with other commonly consumed food legumes (35). Hence, adzuki beans may reduce the risk of breast cancer, although no evidence is available for the mechanism of action of common beans on other kinds of cancer.

A diet high in legume consumption, including adzuki beans, has been associated with a low risk of colorectal cancer (CRC). Adzuki beans contain phytochemicals like phenolic compounds, total dietary fiber (TDF), lectins, unsaturated fatty acids, phytic acid, and trypsin inhibitors, which may prevent or reduce chronic degenerative diseases such as CRC (36–38). Non-digestible fraction (NDF) and phenolic compounds may reach the colon to be fermented by microflora, producing short-chain fatty acids like butyric acid, which can inhibit tumor cell proliferation and induce apoptosis leading to a more differentiated phenotype, thus reducing the risk of CRC (39).

Research has suggested that a hot-water extract of adzuki beans has the potential to stimulate tyrosinase activity in cultured mouse B16 melanoma cells, without inhibiting cell growth (40). This extract is believed to increase the content of cAMP in cells, thereby activating adenyl cyclase and protein kinase pathways and translocating cytosolic PKC to the membrane-bound PKC. The extract has also demonstrated the ability to stimulate hair color pigmentation in C3H mice, indicating potential usefulness in anti-graying and protecting human skin from irradiation. However, these findings only suggest that adzuki bean extract may have pigmentation activity and promote melanogenesis. There is currently insufficient evidence to support the use of adzuki beans as a treatment for melanoma, and further research is warranted.

## C. Dataset Analysis

**Table 2 Final data presentation**

| Network | Food *nutrition | Disease feature | The shape of disease and disease features |
|---|---|---|---|
| NutriFDA | 591*213 | disease-mirna | 320*320 |
| | | disease-gene | 670*670 |
| | | disease ontology | 2109*2109 |
| NutriFDT | 591*213 | disease-mirna | 178*178 |
| | | disease-gene | 208*208 |

|  |  |  |  |  | disease ontology |  | 663*663 |  |
|---|---|---|---|---|---|---|---|---|

Table 2. The result of NutriFDA and NutriFDT networks are presented. Only diseases that are related to food were included in the calculation of disease similarity features. The diseases included in "The shape of disease and disease features" section of the figure represent the intersection between diseases related to food and those with disease feature information.

**Table 3 Results of multi-classification tasks of NutriFDT and NutriFDA**

|  | Disease feature | AUC | F1-score | Model | Max depth | Criterion | No. of categories |
|---|---|---|---|---|---|---|---|
| NutriFDT | Disease-gene | 0.79 | 0.61 | RandomForestClassifier | 2.00 | gini |  |
|  |  | 0.78 | 0.59 | DecisionTreeClassifier | 2.00 | entropy |  |
|  | Disease-miRNA | 0.78 | 0.60 | RandomForestClassifier | 4.00 | gini |  |
|  |  | 0.78 | 0.60 | DecisionTreeClassifier | 2.00 | gini | 12.00 |
|  | DO | 0.79 | 0.61 | RandomForestClassifier | 2.00 | gini |  |
|  |  | 0.79 | 0.61 | DecisionTreeClassifier | 2.00 | gini |  |
| Mean results |  | 0.78 | 0.60 |  |  |  |  |
| NutriFDA | Disease-gene | 0.72 | 0.62 | RandomForestClassifier | 4.00 | gini |  |
|  |  | 0.71 | 0.61 | DecisionTreeClassifier | 2.00 | entropy |  |
|  | Disease-miRNA | 0.70 | 0.59 | RandomForestClassifier | 4.00 | gini |  |
|  |  | 0.69 | 0.59 | DecisionTreeClassifier | 2.00 | gini | 3.00 |
|  | DO | 0.71 | 0.61 | RandomForestClassifier | 2.00 | gini |  |
|  |  | 0.69 | 0.58 | DecisionTreeClassifier | 2.00 | gini |  |
| Mean results |  | 0.70 | 0.60 |  |  |  |  |

Table 3. The value of 0 was classified as a separate category, while the remaining categories were grouped using a binning method.

The results of multi-classification models on NutriFDT are shown in Table 3. The area under the receiver operating characteristic curve (AUC) and F1 Score are robust evaluation metrics that work great for many classification problems. F1 score combines precision and recalls into one metric by calculating the harmonic mean (41) . The AUC is a widely-used quantitative measure of classification performance (42). The higher the value, the better the classifier effect.

The results of the study in Table 3 indicate that both the random forest and decision tree models demonstrated high performance for NutriFDT and NutriFDA, suggesting that these networks possess strong generalization abilities. However, due to the decrease in performance observed with an increase in category types in NutriFDA, we restricted the categories of NutriFDA to three types. This situation may be explained by the fact that nutrition can play a therapeutic role in dietary treatments, rather than simply displaying association relationships with diseases.

## Discussion

Nutritional components of food interact with specific protein genes indirectly or directly to influence the development of diseases or even prevent their occurrence (43, 9). To observe the relationship between food and disease from the perspective of nutrients, this paper made certain assumptions about the compound matching method, food matching method, and inference scores of the compound-disease relationship, and established the "nutrient-food-disease-disease similarity feature" connection. The number of associations between food and disease reaches as 3237640, while the amount of treatment relation between those reaches as 170531. Previously, the study of the relationship between food and disease from a nutritional perspective had focused on experimental studies, while this paper validated dietary feasibility of nutrition from a perspective of big data. To verify the advanced nature of NutriFDA and NutriFDT, it is compared and analyzed with NutriChem, FooDisNET, and ABCkb in four aspects.

1) **The amount of data in the food-disease dataset**

The following table compares the amount of connected data between NutriFDA, NutriFDT, and other food-disease databases.

**Table 4 Food-disease associations in different networks**

| Disease similarity type | Food | Disease | Food-disease association |
|---|---|---|---|
| NutriFDA[*#] | 591 | 2208 | 3237640 |
| NutriFDT[*#] | 591 | 671 | 170531 |
| NutriChem (9) [*#] | 1772 | 7828 | |
| FooDisNET (10) [*] | 6329 | 18,689 | 53,920 |
| ABCkb (19) [#] | 229143 | 53302 | 29,124 |

Table 4. Note that the "*" symbol indicates that the database primarily focuses on associations, while the "#" symbol indicates that the database focuses on treatment relations.

The results, given in Table 4, show that in addition to exploring more food-disease relationships, NutriFD focuses on everyday foods and provides a new paradigm for evaluating therapeutic relationships, which haven't existed in other databases.

2) **Compound integration method**

In this paper, compounds were expressed using InChiKey, which is a more unique and complete method of compound expression compared to SMILES used in NutriChem. Compared with DisGeNET (44) and ABCkb (19) databases that expressed compounds with InChiKey, this paper incorporated a multi-layer check of compound name, mesh, and CASRN in the compound matching process, which largely improved the accuracy of the database.

**3） Accession of food and disease characteristics**

The database built in this paper incorporated food nutrient element characteristics, which were not considered in previous food-disease databases. Besides, we successfully proved that nutrients dominate the effect of dietary therapy, allowing machine learning models to generalize well.

**4） Food-disease scoring method**

This paper utilized the inference score data calculated by Common Neighbor Statistics (CNS) in CTD (22,44), which takes into account the compound-gene-disease local network topology information, i.e. when two similar nodes share many interacting neighbors, the local network topology information can significantly increase the possibility of two nodes sharing related biological functions. Compared with the food-disease association scoring method defined by FooDisNET from four perspectives of compound sparsity, protein universality, disease-protein association, and food compound association, the food-disease scoring relation established in this paper has more biological significance.

## New Perspectives

Although there are challenges in conducting diet trials (29), the use of nutrition to treat disease represents a unique opportunity. Current diet research is focused on designing and evaluating effective interventions, with less research on clear molecular pathway targets for diet therapy. This dataset might provide researchers with a food-related dataset for disease, thus providing guidance for researchers to explore the direction of trials and helping to guide future trials of dietary interventions.

We also provide food nutrient profiles from USDA and diseases similarity profiles for researchers to study the potential relationship between them. Researchers can propose improved machine learning models to enhance the accuracy of predicting the relationship between food and disease. it is also hoped that researchers can use KEGG (45,46) to establish an inference score that takes into account the metabolic pathways of multiple compounds.

## Conclusion

Although food therapy has long been developed in the field of nutrition to analyze nutritional elements and meet different nutritional needs during the human life cycle, the establishment of the relationship between food and disease from the perspective of big data is a recent emergence. In this study, we developed NutriFDA, an association scoring network, and NutriFDT, a treatment scoring network. Furthermore, we utilized machine learning models to evaluate the predictive nature of nutrition. This dataset may provide researchers with food-related datasets for diseases, which can inform trials of dietary therapy and guide future interventions.

## Data availability

The authors declare that the NutriFD knowledgebase supporting the findings of this study is available via https://nutrifd.dedyn.io. The following databases are used in this study: USDA (https://fdc.nal.usda.gov/download-datasets.html), CTD (http://ctdbase.org/downloads/), FooDB (https://foodb.ca/downloads), HMDD v3.2 (https://www.cuilab.cn/hmdd), miRCancer (http://mircancer.ecu.edu/), DisGeNET (https://www.disgenet.org/downloads#), SNAP (https://snap.stanford.edu/biodata/), NIH (https://disease-ontology.org/downloads/), TXT data for disease-gene associations [47].

## Acknowledgements

This work has been supported by the CAS Light of the West Multidisciplinary Team project [xbzg-zdsys-202114], and the Pioneer Hundred Talents Program of Chinese Academy of Sciences.

## Author contributions

P.H. conceived and designed the work. W.S. was responsible for acquiring, harmonizing, integrating, and evaluating the databases, as well as discovering and approving case studies. P.H., W.S. and L.H. co-wrote the manuscript. D.L. and F.T. contributed to the design of the platform.

## Competing interests

The authors declare no competing interests.